\documentclass[12pt]{iopart}
\input epsf.tex
\eqnobysec

\newcommand{\bsy}[1]{\mbox{\boldmath $#1$}}

\def\be{\begin{equation}}
\def\ee{\end{equation}}

\begin{document}
\jl{1}

\title{Gradient critical phenomena in the Ising quantum chain}

\author{T. Platini,
	D. Karevski
	\footnote[1]{Corresponding author: {\tt karevski@lpm.u-nancy.fr}}
	and 
	L. Turban
}
\address{Laboratoire de Physique des Mat\'eriaux, UMR CNRS 7556,\\ 
  Universit\'e Henri Poincar\'e, Nancy 1,\\ 
  B.P. 239,  F-54506  Vand\oe uvre l\`es Nancy Cedex, France
}

\begin{abstract}
We consider the behaviour of a critical system in the presence of a gradient perturbation of the couplings. In the direction of the gradient  an interface region separates the ordered phase from the disordered one. We develop a scaling theory for the 
density  profiles induced by the gradient perturbation which involves a characteristic length given by the width of the interface region. The scaling predictions are tested in the framework of the mean-field  Ginzburg-Landau theory. Then we consider the Ising quantum chain in a linearly varying transverse field which corresponds to the extreme anisotropic limit of a classical two-dimensional Ising model. The quantum Hamiltonian can be diagonalized  exactly in the scaling limit where the eigenvalue problem is the same as for the quantum harmonic oscillator. The energy density, the  magnetization profile and the two-point correlation function are studied either analytically or by exact numerical calculations. Their scaling behaviour  are in  agreement with the predictions of the scaling theory.
\end{abstract}

\section{Introduction}
Inhomogeneities may have a strong influence  on the properties of a system in the vicinity of a second-order phase transition. Actually this influence will depend  on the relevance of the perturbation introduced by the inhomogeneity (see~\cite{igloi93} for a review). A relevant  inhomogeneity may change the universality class of the system  or even suppress the critical point as, for instance, in  a finite-size system~\cite{binder83}. The  critical behaviour may be altered by  the presence of quenched disorder~\cite{berche04} or  aperiodic modulation of the couplings~\cite{luck93,grimm97}. It will be modified  locally (i.e., within a  correlation length) at a flat free surface, at a corner~[6--12]
 or at the tip of a parabolic-shape system~\cite{peschel91,igloi93}. Line defects may have also some influence on the  
 local critical behaviour~[14--16,1]. One may  mention the case of films where the presence 
 of boundaries, by  breaking  translation invariance, leads to
 the formation of specific profiles~[17--24].

Other types of inhomogeneities are linked to the application of  external fields  like magnetic,  gravitational  
or thermal  fields. These fields influence the behaviour of physical quantities like the magnetization,  the particle  or the energy density in the vicinity of the homogeneous system  critical point.  One should first mention an early work  on  the $xy$-quantum chain in a linearly varying $z$-field~\cite{smith71} for which exact results were obtained in different scaling limits.  Phase coexistence induced by antiparallel magnetic  fields at the surfaces of a film was studied in~[26--30]. The effect of gravity was  considered in~[28--30]  where it was found that it restores two-phase coexistence up to the bulk critical point, above the wetting temperature.  The effect of a temperature gradient  on an interface at equilibrium was considered in~\cite{boyanovsky95} for  a symmetric binary system below its critical point. Phase separation induced by temperature gradients was studied in~[32--34].The effect of temperature gradients on interfacial premelting was considered in~\cite{rempel01}. 
 
 The influence of  inhomogeneities  on the critical behaviour  was also considered in a series of works on gradient 
 percolation~[36--40]  where the  perturbation was introduced as a tool allowing 
 for high-precision  estimates  of the percolation threshold and  the percolation exponents. 
 
In this work  we begin with a presentation of  the scaling theory  for the density profiles in the presence of gradient field inhomogeneities.  Specifically, we consider a system with a deviation from the critical coupling which varies  linearly in one space direction. The coupling is at its critical value in the middle of the system where an interface  region separates the ordered phase on the left  from the disordered phase on the right.  We test the validity of the scaling arguments, first at the mean-field level, within  Ginsburg-Landau theory. Then we present a study of the Ising quantum chain in a linearly varying transverse field, $h_l=1+gl$,  which corresponds to the extreme anisotropic limit  of  the two-dimensional  classical Ising model with a linear variation of the couplings. We work in the scaling limit where the size $L$ of the system goes to infinity while the gradient $g$ goes to zero with the product $gL$ held fixed.  The excitation spectrum of  the inhomogeneous Ising quantum chain is obtained exactly  in terms of the solution of  an harmonic oscillator eigenvalue problem. The  knowledge of the eigenvectors allows us to obtain the  energy  density profile, the magnetization profile  and the behaviour of the spin-spin correlation function. Their scaling forms are in complete  agreement with the results of the scaling theory.

The paper is organized as follows: in  section~2  we present a  scaling analysis for the density profiles which is confirmed in section~3 in the framework of  a  mean-field approximation. Section~4 deals with the study the Ising quantum  in a linearly varying transverse field. First  the form of the quantum Hamiltonian is deduced from the classical Ising problem. Then the Hamiltonian is diagonalized exactly  in the scaling limit. In the next subsections, analytical and  numerical results for the energy density, the magnetization profile  and the correlation function  are confronted to the results of the scaling theory. We summarize  our results in the last section.

\section{Scaling arguments}
Let us  consider  a critical system perturbed by a constant   gradient $g$  along the $z$-direction,  such that the deviation from the critical coupling  is given by
\be
K(z)-K_{\rm c}=\Delta(z)=gz\,,\qquad g>0\,.
\label{2-1}
\ee
Thus around the origin  there is an interface between the  ordered phase on the left side  where $\Delta<0$  and the disordered phase on the right side where  $\Delta>0$. Let $\ell$ denote the width of this interface. If the system is infinite $\ell$ is expected to depend on $g$  only and to diverge when $g$ vanishes.

Under a change of the length scale by a factor $b$, the thermal perturbation $\Delta(z)$ with scaling dimension $y_t=1/\nu$, where $\nu$ is the correlation length exponent, transforms as
\be
g'z'=b^{1/\nu}gz=g'\frac{z}{b}\,,
\label{2-2}
\ee
so that 
\be
g'=b^{1+1/\nu}g\,.
\label{2-3}
\ee
The interface width transforms as 
\be 
{\ell}'=\ell(g')=\frac{\ell}{b}=\ell\left(b^{1+1/\nu}g\right)\,,
\label{2-4}
\ee
with $b=g^{-\nu/(1+\nu)}$, one finally obtains
\be
\ell\propto g^{-\nu/(1+\nu)}\,,
\label{2-5}
\ee
for the typical length introduced by the thermal  gradient $g$.

The same result can be obtained self-consistantly~\cite{sapoval85} by noticing that to the width $\ell$ is associated a typical deviation from the critical coupling  $\Delta(\ell)=g\ell$ from which a  characteristic  length $[\Delta(\ell)]^{-\nu}$ can be constructed. Since the only   length in the problem is  the interface width, $\ell$, it  satisfies
\be
\ell\propto (g\ell)^{-\nu}\,.
\label{2-6}
\ee
Solving for $\ell$, one immediately recovers~\eref{2-5}.
 
Let us now study the influence of the thermal  gradient on the scaling  behaviour of the density $\varphi$ with scaling dimension $x_\varphi$. This density can be the magnetization density $m$ or the singular part of the energy density $e$. In the perturbed critical system it is a function $\varphi(z,\ell)$ or, alternatively, according to~\eref{2-5}, a function $\varphi(z,g)$ transforming as 
\be
{\varphi}'=\varphi(z',g')=b^{x_\varphi}\varphi(z,g)
\label{2-7}
\ee
under a change of scale, so that
\be
\varphi(z,g)=b^{-x_\varphi}\varphi\left(\frac{z}{b},b^{1+1/\nu}g\right)\,.
\label{2-8}
\ee
With $b=g^{-\nu/(1+\nu)}\propto\ell$ one obtains the scaling form
\be
\varphi(z,g)=g^{\nu x_\varphi/(1+\nu)}\Phi \left[g^{\nu/(1+\nu)} z\right]\, .
\label{2-9}
\ee
One may notice that, according to~\eref{2-5}, the prefactor  in~\eref{2-9} exhibits the finite-size behaviour $\varphi\propto\ell^{-x_\varphi}$  expected for a critical system with transverse size $\ell$.

For a magnetic system the magnetization density, $m(z,g)$, with scaling dimension $x_{\rm m}=\beta/\nu$,  is non-vanishing in the ordered region $z<0$ and one expects the same local critical behaviour as for the homogeneous system at the same value of the coupling, i.e.,
\be
m(z,g)\propto|\Delta(z)|^{\beta}\propto\left|gz\right|^{\nu x_{\rm m}}\,,\qquad z<0\,.
\label{2-10}
\ee
for not too large values of $|\Delta(z)|$. Here and in the following, we suppose that $m(z,g)\geq0$. Comparing~\eref{2-10} to~\eref{2-9} with $\varphi=m$, one obtains the form of the scaling function in this region:
\be
\Phi_{\rm m}(u) \sim |u|^{\nu x_{\rm m}}\,,\qquad u<0\,.
\label{2-11}
\ee
Just at the interface one expects $\Phi_{\rm m}(0)={\rm const.\neq 0}$ so that the local magnetization is non-vanishing and behaves as the prefactor in~\eref{2-9}, displaying the finite-size behaviour $m(0,g)\propto\ell^{-x_{\rm m}}$. Since $m(z,g)>m(0,g)$ when $z<0$, the scaling behaviour given in equations~\eref{2-10} and~\eref{2-11} is valid  only when $\left|gz\right|^{\nu x_{\rm m}}>\ell^{-x_{\rm m}}$. Using~\eref{2-5}, this translates into $z<-\ell$.

On the right of the interface, in the weak-coupling region, the magnetization density is expected to display the same exponential decay as   the two-point correlation function $\Gamma(z)=\langle m(0)m(z)\rangle$. For a homogeneous system in its disordered phase, with a constant deviation $\Delta$ from the critical coupling, one has
\be
\Gamma(z)\sim \exp\left(-{\rm const.}\, \frac{z} {\xi}\right)\sim \exp\left(-{\rm const.}\, \frac{z} {\Delta^{-\nu}}\right)\,,\qquad z\gg\xi\,,
\label{2-12}
\ee
where $\xi$ is the correlation length. Assuming  the same behaviour for the inhomogeneous system, with the correlation length taking a value governed by the local deviation from the critical point, $\xi\propto[\Delta(z)]^{-\nu}$, one obtains~\cite{peschel91}
\be
m(z,g)\sim \exp\left(-{\rm const.}\, g^\nu z^{1+\nu}\right)\,.
\label{2-13}
\ee
Equations~\eref{2-9} and~\eref{2-13} give the form of  the scaling function for  the order parameter when $z\gg0$ 
\be
m(z,g)=g^{\nu x_{\rm m}/(1+\nu)}\Phi_{\rm m} \left[g^{\nu/(1+\nu)} z\right],\quad\Phi_{\rm m}(u) \sim \exp \left(-{\rm const.}\, u^{1+\nu}\right).
\label{2-14}
\ee

The same type of arguments leads to the scaling behaviour of  the singular part of the energy density
\be
e(z,g)=g^{\nu x_{\rm e}/(1+\nu)}\Phi_{\rm e} \left[g^{\nu/(1+\nu)} z\right]\,,\quad \Phi_{\rm e}(u) \sim \exp \left(-{\rm const.}\, |u|^{1+\nu}\right)\,,
\label{2-15}
\ee
for both sides of the interface. Here $x_{\rm e}=d-1/\nu$ is the scaling dimension of the energy density.
 
The two-point correlation function, with scaling dimension $2x_{\rm m}$, can also be written under the scaling form
\be
\Gamma(z,g)=g^{2\nu x_{\rm m}/(1+\nu)}\Phi_{\Gamma} \left[g^{\nu/(1+\nu)} z\right]\,,
\label{2-16}
\ee
with the exponential decay, $\Phi_{\Gamma}(u) \sim \exp \left(-{\rm const.}\, u^{1+\nu}\right)$, for  the connected part, $\Gamma_{\rm c}(z,g)=\Gamma(z,g)-\langle m(0)\rangle\langle m(z)\rangle$.

\section{Mean-field theory}
In order to check our scaling assumptions, let us now study the gradient perturbation problem in mean-fied theory. With a scalar order parameter $m$ and up-down symmetry,  the Ginzburg-Landau free energy functional of the critical system perturbed by the gradient term reads
\be
G[m]=G[0]+\int_V \left[\frac{C}{2}( \bsy{\nabla} m)^2+\frac{\Delta(z)}{2}m^2+\frac{B}{4}m^4-Hm\right]dV\,,
\label{3-1}
\ee
where $B$ and $C$ are positive constants. The first term is the energy contribution coming from inhomogeneities. The quadratic  term is the thermal gradient perturbation with, as before, $\Delta(z)=gz$ and $g>0$. The quartic term ensures the stability of the system in the ordered region for $z<0$. The last term gives the interaction with the external field $H$. 

In the mean-field approximation the equilibrium value of the order parameter minimizes $G[m]$. Thus the variation of the free energy, $\delta G$, vanishes to first order in $\delta m$. This leads to the  Ginzburg-Landau equation
\be
-C\frac{d^2 m}{d z^2}+gz\, m(z)+B\, m^3(z)=0\,,
\label{3-2}
\ee
where it was assumed that $H=0$ and that translation invariance is broken only in the $z$-direction.  

Introducing the dimensionless variable, $\zeta=z/\ell$,  the Ginzburg-Landau equation can be rewritten as
\be
-\frac{C}{\ell^2}\frac{d^2 m}{d \zeta^2}+g\ell\zeta m(\zeta)+Bm^3(\zeta)=0\,.
\label{3-3}
\ee
The coefficients of  $d^2 m/d \zeta^2$ and $m(\zeta)$ have to scale in the same way.  Thus 
$C/\ell^2\propto g\ell$ and the  interface width has the following scaling behaviour
\be
\ell\simeq\left(\frac{C}{g}\right)^{1/3}\propto g^{-\nu/(1+\nu)}\,,
\label{3-4}
\ee
since $\nu=1/2$ in mean-field theory.

In the ordered region, for $z\ll -\ell$ or $\zeta\ll -1$ in~\eref{3-3}, one can neglect the second derivative which is much smaller than the next term in the Ginzburg-Landau equation~\eref{3-2} so that we obtain
\be
m(z,g)\simeq \left(\frac{-gz}{B}\right)^{1/2}\,,
\label{3-5}
\ee
which confirms the scaling prediction~\eref{2-10} since here $\beta=\nu x_{\rm m}=1/2$.

The order parameter is very small for $z\gg0$ and the cubic term in~\eref{3-2} can be neglected. With the change of variable $u=z(g/C)^{1/3}$ one obtains the Airy equation
\be
\frac{d^2 m}{d u^2}-u\, m(u)=0\,.
\label{3-6}
\ee
The order parameter  is then given by
$m(z,g)\propto{\rm Ai}(u)$ with the leading asymptotic behaviour
\be
m(z,g)\sim \exp\left(-{\rm const.} \,u^{3/2}\right)\sim\exp(-{\rm const.} \,g^{1/2} z^{3/2})\,,
\label{3-7}
\ee
when $z\gg0$ in agreement with equation~\eref{2-14} with $\nu=1/2$.

\section{Ising quantum chain in a linearly varying transverse field}
\subsection{Quantum Hamiltonian}
Let us consider the two-dimensional classical nearest-neighbour Ising model on a square lattice with vertical couplings $K_1(l)$  varying in the horizontal direction ($-L/2<l<L/2$) and constant  horizontal   couplings $K_2$. The partition function can be written as
\be
Z\propto{\rm Tr}{\cal T}^M\,,
\label{4-1}
\ee
where $M$ is the number of horizontal rows and ${\cal T}$ is the row-to-row transfer matrix given by~\cite{onsager44,schultz64}
\be
{\cal T}=\exp\left[\sum_l K_1^*(l)\,\sigma_l^z\right]
\exp\left[\sum_lK_2\,\sigma_l^x\sigma_{l+1}^x\right]\,.
\label{4-2}
\ee
The $\sigma^{x,z}$ are Pauli spin operators, $K_1^*(l)=-1/2\ln\tanh[K_1(l)]$ is the dual of the vertical coupling and it  is assumed to vary as
\be
K_1^*(l)= K_1^*(1+gl)\,,\qquad g\geq 0\,.
\label{4-3}
\ee

In the extreme anisotropic limit~[43--45] $K_1\to \infty$ (so that  $K_1^*\to 0$)  and $K_2\to 0$ while keeping the ratio $h=K_1^*/K_2$ constant,  the transfert matrix $\cal T$ can be rewritten as
\be
{\cal T}=1-2K_2{\cal H}\,.
\label{4-4}
\ee
${\cal H}$ is the Hamiltonian of the Ising quantum chain in a transverse field~\cite{pfeuty70} and takes the following form
\be
{\cal H}=-\frac{1}{2}\sum_{l=-L/2}^{L/2-1}\sigma^x_l\sigma^x_{l+1}
-\frac{1}{2}\sum_{l=-L/2}^{L/2}h_l\,\sigma^z_l\,, \qquad h_l=h(1+gl)\,,
\label{4-5}
\ee
where $g=\theta/L$ with $\theta\geq 0$.

In the homogeneous case, $g=0$, the system is  self-dual and critical at $h=1$ in the thermodynamic limit $L\to\infty$. For $h<1$ the system is ordered with $\langle \sigma^x\rangle\ne 0$ whereas $\langle \sigma^x\rangle=0$ when $h>1$.

The perturbed system is assumed to be critical when $\theta=0$ so that  $h=1$ in~\eref{4-5} and $h_l=1+\theta l/L=1+gl$. In order to keep $h_l\geq 0$, one has to take $\theta\leq 2$. The transverse field is smaller (greater) than its critical value on the left (right) of the origin. Thus, in the thermodynamic limit, the quantum chain is ordered for $l<0$  and the value of the order parameter decreases to zero when $l\to+\infty$. 

\subsection{Diagonalization}
After a Jordan-Wigner transformation~\cite{jordan28}, the
Hamiltonian~\eref{4-5} becomes a quadratic form in fermion creation and
annihilation operators which is diagonalized through a canonical
transformation~\cite{pfeuty70,lieb61} leading to
\be 
{\cal H}= \sum_{q=0}^{L}\varepsilon_{q}(\eta_{q}^{\dag}\eta_{q}-\frac{1}{2})\,.
\label{4-6}
\ee
The $\eta_{q}^{\dag}$  ($\eta_{q}$ ) are diagonal fermion creation
(anihilation) operators and the $\varepsilon_q$  are the energies of
the fermionic excitations. They satisfy the following set of equations
\be
{\mathsf A}\psi_q=\varepsilon_q\phi_q\,,\qquad{\mathsf A}^\dag\phi_q=\varepsilon_q\psi_q\,,
\label{4-7}
\ee
where 
\begin{eqnarray}
{\mathsf A}=\left(\begin{array}{llllll}
-h_{-L/2}&0	&0	&0	&0\\
1      &-h_{-L/2+1}&0	&0&0 \\
	&  \ddots& \ddots&& \\
0	&0		& 1	&-h_{L/2-1}&0\\
0	&0	&0	& 1	&-h_{L/2}\\
\end{array}\right)
\label{4-8}
\end{eqnarray}
and ${\mathsf A}^\dag$ is the transposed matrix. According to \eref{4-7}, the normalized eigenvectors, $\phi_q$ and $\psi_q$, are solutions of the following eigenvalue equations
\be
 {\mathsf A}{\mathsf A}^\dag\phi_q=\varepsilon_q^2\phi_q\,,\qquad 
 {\mathsf A}^\dag{\mathsf A}\psi_q=\varepsilon_q^2\psi_q\,.
 \label{4-9}
 \ee
In the bulk of the system  these  eigenvalue equations  can be  written as
\begin{eqnarray}
&&h_{l-1}\,\phi_q(l-1)+\left[\varepsilon_q^2-1-h_l^2\right]\phi_q(l)+h_{l}\,\phi_q(l+1)=0\,,
 \nonumber\\
&&h_{l}\,\psi_q(l-1)+\left[\varepsilon_q^2-1-h_l^2\right]\psi_q(l)+h_{l+1}\,\psi_q(l+1)=0\,.
\label{4-10}
\end{eqnarray}

Introducing the scaling variable $u=g^{1/2}l$, the difference equations in~\eref{4-10} can be expanded in the scaling  limit  where $L\to\infty$ and $g\to0$ in such a way that the  product, $gL=\theta$,  is held fixed. Up to terms of the first order in $g$ the expansions lead  to the following  harmonic oscillator eigenvalue  equations
\be\fl
\frac{d^2\phi }{d u^2}+\left[\left(\frac{\varepsilon}{g^{1/2}}\right)^2 -1-u^2 \right]\phi(u)=0\,,\qquad
\frac{d^2\psi }{d u^2}+\left[\left(\frac{\varepsilon}{g^{1/2}}\right)^2+1-u^2 \right]\psi(u)=0\,.
\label{4-11}
\ee
With the boundary conditions $\phi(\pm \infty)=\psi(\pm \infty)=0$, one obtains
\be\fl
\psi_n(u)=C_n{\rm e}^{-u^2/2}H_n(u)\,, \quad \phi_{n+1}(u)=\psi_n(u)\,, \quad \varepsilon_n=\sqrt{2ng}\,,\quad n=0,1,2,\dots\,,
\label{4-12}
\ee
where $H_n(u)$ is the Hermite polynomial of order $n$ and  $C_n=(g/\pi)^{1/4}(2^n n!)^{-1/2}$ is a normalization factor. 

One may notice that in the bulk, according to ~\eref{4-8},  the operators ${\mathsf A}$ and ${\mathsf A}^\dag$ satisfy  the  commutation relation:
\be
[{\mathsf A},{\mathsf A}^\dag]=g{\mathsf C}\,,\qquad {\mathsf C}_{l,m}=\delta_{l-1,m}+\delta_{l+1,m}\,.
\label{4-13}
\ee 
Introducing the normalized operators, ${\mathsf a}={\mathsf A}/\sqrt{2g}$ and ${\mathsf a}^\dag={\mathsf A}^\dag/\sqrt{2g}$, in~\eref{4-13} and applying the commutator to a test function, in the scaling limit one obtains the canonical bosonic commutation relation
\be
[{\mathsf a},{\mathsf a}^\dag]=1+\Or(g)\,.
\label{4-14}
\ee
Thus, once normalized in the same way, the equations in~\eref{4-7} correspond to the lowering and raising operations on the eigenstates $\psi_n$ and $\phi_n$ of the harmonic oscillator.

The  eigenstate $\phi_0(u)$, associated with the  excitation $\varepsilon_0$ which is vanishing in the scaling limit, cannot be obtained from the harmonic oscillator equation since it is incompatible with the boundary condition at $-\infty$. It corresponds to a  mode which is localized in the vicinity of the left boundary and which is related to the presence of a non-vanishing  magnetization $m_{\rm s}$ at the left boundary of the system~\cite{peschel84,karevski00}. In order to obtain the form of this localized mode and the actual value of the corresponding  excitation, one has to come back to the original  finite-size system with open boundary conditions. It can be shown that,  when it  vanihishes faster than $1/L$,   the lowest excitation is given by~[52--54]
\be
\varepsilon_0 \simeq m_{\rm s} m^*_{\rm s}\,,
\label{4-15}
\ee
where $m^*_{\rm s}$ is the magnetization of the dual chain at the left boundary.
 $m_{\rm s}$ and $m^*_{\rm s}$ are  equal to the component $\phi_0(-L/2)$ of the normalized eigenvector 
 and can be deduced from the second equation in~\eref{4-7} with $\varepsilon_0=0$. 
 These  boundary magnetizations take the following  forms~\cite{peschel84}
\be
\fl
m_{\rm s}=\left[1+\sum_{k=-L/2}^{L/2}\prod_{l=-L/2}^{k}h_l^2\right]^{-1/2}\,,\qquad
m^*_{\rm s}=\left[1+\sum_{k=-L/2}^{L/2}\prod_{l=-L/2}^{k}h_l^{-2}\right]^{-1/2}\,.
\label{4-16}
\ee
With $h_l=1+gl$  it is straightforward to show that $m_{\rm s}$  is $\Or(1)$  whereas  the dual magnetization is exponentially small:
\be 
 m^*_{\rm s}\simeq \left(\frac{g}{\pi}\right)^{1/4}\exp\left(-\frac{g L^2}{8}\right)\,.
\label{4-17}
\ee
As a consequence,  the smallest excitation vanishes exponentially  with the size of the system as
\be
\varepsilon_0\sim \e^{-gL^2/8}\,.
\label{4-18}
\ee
Using the second equation in~(\ref{4-13}) with $\varepsilon_0\sim 0$, one obtains 
\be
\phi_0(-L/2+l)\simeq\phi_0(-L/2)\e^{- l/l_0}\,,\qquad l_0^{-1}=|\ln(1-gL/2)|\,.
\label{4-19}
\ee
The localisation length $l_0$ diverges as $(\theta/2)^{-1}$ when the  local deviation from 
the critical transverse field, $\theta/2=1-h_{-L/2}$, vanishes. It behaves as the local correlation length since $\nu=1$ for the Ising quantum chain.
In the scaling limit  $g\to0$ and $L\to\infty$ while $\theta=gL$ remains constant;  thus the localization 
length $l_0$ remains also constant  and the localized mode, $\phi_0(u)$, has a vanishing amplitude in the bulk.
 
This completes the diagonalization of the model in the continuum limit.

%%%%%%%%%%%%%%%%%%%%%%%%% fig 1 %%%%%%%%%%%%%%%%%%%%%%%%
\begin{figure} [tbh]
%\vspace{0.2cm}
        \epsfxsize=9cm
        \begin{center}
        \mbox{\epsfbox{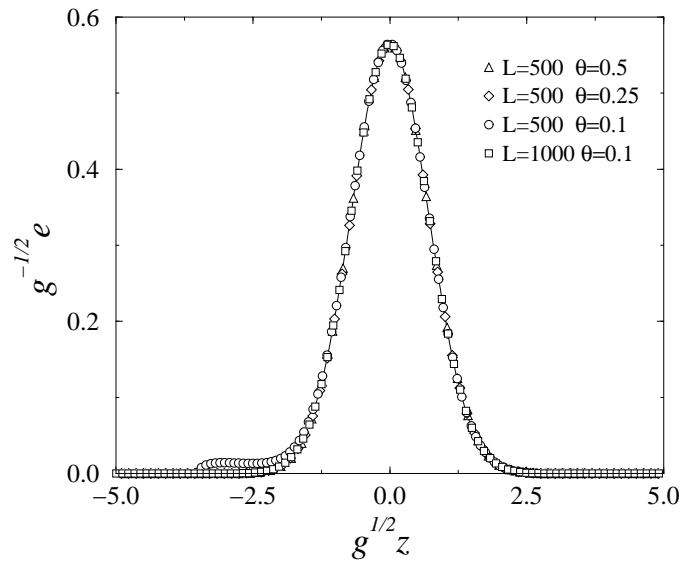}}
        \end{center} 
	\vskip -.8cm
        \caption{Rescaled energy density profiles for different chain sizes $L$ and  $\theta$ values. 
	The corresponding values of the gradient are $g=1\times 10^{-3}, 5\times 10^{-4}, 2\times 10^{-4}, 1\times 10^{-4}$ 
	from top to bottom in the legend.
	The solid line is  the analytic result in~ \protect\eref{4-24}. }
        \label{Fig1}  \vskip 0cm
\end{figure}
%%%%%%%%%%%%%%%%%%%%%%%%%%%%%%%%%%%%%%%%%%%%%%%%%%%%

\subsection{Energy density profile}
Let us consider the connected autocorrelation function in imaginary time
\be
G_{\rm e}(l,\tau)=\langle\sigma_l^z(\tau)\sigma_l^z\rangle-\langle\sigma_l^z\rangle^2\,,
\label{4-20}
\ee
where $\sigma_l^z(\tau)={\rm e}^{\tau {\cal H}}\sigma_l^z{\rm e}^{-\tau {\cal H}}$.
It can be rewritten as the eigenstate expansion
\be
G_{\rm e}(l,\tau)=\sum_{i>0}|\langle i|\sigma_l^z|0\rangle|^2 \e^{-\tau(E_i-E_0)}\,,
\label{4-21}
\ee
where $|0\rangle$ is the ground state of ${\cal H}$ with eigenvalue $E_0$ and $|i\rangle$  an excited state  with eigenvalue $E_i$.
Since the operator $\sigma^z_l$ is a two-fermion operator, the only non-vanishing matrix elements in the  expansion are the two-fermion states $|i\rangle=\eta_q^\dag\eta_p^\dag|0\rangle$. When $\tau\rightarrow \infty$, the amplitude of the dominant term defines the off-diagonal energy density
\be
e(l)=|\langle \epsilon|\sigma_l^z|0\rangle|\,,
\label{4-22}
\ee
where  $|\epsilon\rangle =\eta_1^\dag\eta_0^\dag|0\rangle$ is the first even excited state. The expansion  of $\sigma_l^z$ in terms of diagonal fermions leads to~\cite{berche90}
\be
e(l)=|\psi_1(l)\phi_0(l)-\psi_0(l)\phi_1(l)|\,.
\label{4-23}
\ee
Inserting~(\ref{4-12}) into~(\ref{4-23}), one finally obtains  the Gaussian profile
\be
e(l)=|\psi_0(l)|^2=\sqrt{\frac{g}{\pi }}\exp(-g l^2) \,,
\label{4-24}
\ee
which is in agreement with the form~\eref{2-15} deduced from scaling considerations  since $\nu=x_{\rm e}=1$ in the Ising model.

In order to reduce the  boundary effects due to the localized mode, one has to consider sizes $L\gg l_0\sim\theta^{-1}$ for small $\theta$. Typically, for $\theta=0.1$, $l_0\simeq 20$ and taking $L>10\times l_0$ is sufficient to suppress the localized mode effect. The behaviour of the off-diagonal  energy density,  deduced from \eref{4-23} through  numerically  exact diagonalization for chains of size  up to $L=1000$,  is shown in  figure~\ref{Fig1}. An excellent agreement with the analytic result in~\eref{4-24} is obtained.
%%%%%%%%%%%%%%%%%%%%%%%%% fig 2 %%%%%%%%%%%%%%%%%%%%%%%%
\begin{figure} [tbh]
%\vspace{0.2cm}
        \epsfxsize=9cm
        \begin{center}
        \mbox{\epsfbox{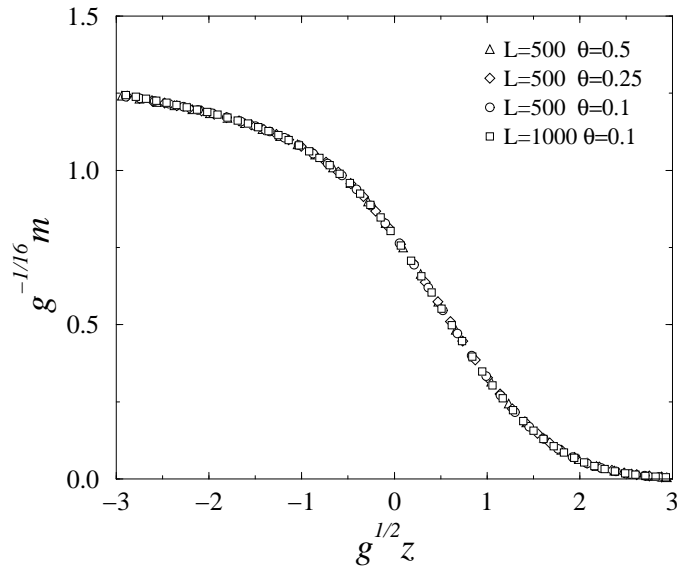}}
        \end{center}
	\vskip -.8cm
        \caption{Rescaled magnetization  profiles for different chain sizes $L$ and  $\theta$ values. 
	The values of the gradient are the same as in figure~\ref{Fig1}.}
        \label{Fig2}
\end{figure}
%%%%%%%%%%%%%%%%%%%%%%%%%%%%%%%%%%%%%%%%%%%%%%%%%%%

\subsection{Magnetization profile}
Due to the $Z_2$ symmetry of the Hamiltonian, the ground state expectation value of 
$\sigma_l^x$ identically vanishes. In order to obtain the magnetization profile one has to break this symmetry. Since the ground state $|0\rangle$ is asymptotically degenerate with the lowest one-fermion state $|\sigma\rangle=\eta_0^+|0\rangle$ such  that
$E_\sigma-E_0=\varepsilon_0\simeq 0$, one can construct the state
\be
|+\rangle =\frac{1}{\sqrt{2}}\left(|0\rangle+|\sigma\rangle\right)
\label{4-25}
\ee
and take the expectation value of $\sigma_l^x$ in that state which leads to\footnote[2]{More generally, with the linear combination  $|\alpha\rangle=\frac{|0\rangle+\alpha|\sigma\rangle}{\sqrt{1+\rho^2}}$,  where $\alpha=\rho{\rm e}^{i\theta}$, one obtains 
$\langle\alpha|\sigma_l^x|\alpha\rangle=  \frac{2\rho\cos\theta}{{1+\rho^2}} \langle \sigma|\sigma_l^x|0\rangle$ which is maximum   for $\alpha=1$ when $\langle \alpha|\sigma_l^x|\psi_\alpha\rangle= \langle \sigma|\sigma_l^x|0\rangle$.} 
\be
m(l)=\langle +|\sigma_l^x|+\rangle=\langle \sigma|\sigma_l^x|0\rangle\,.
\label{4-26}
\ee
One may also notice that  the imaginary time autocorrelation function has the following asymptotic behaviour
\be
\lim_{\tau\to\infty}G_m(l,\tau)=m^2(l)=\langle \sigma|\sigma_l^x|0\rangle^2\,,
\label{4-27}
\ee
which is another way to justify the use of the off-diagonal matrix element  for the magnetization profile.
 
%%%%%%%%%%%%%%%%%%%%%%%%% fig 3 %%%%%%%%%%%%%%%%%%%%%%%%
\begin{figure} [tbh]
        \epsfxsize=9cm
        \begin{center}
        \mbox{\epsfbox{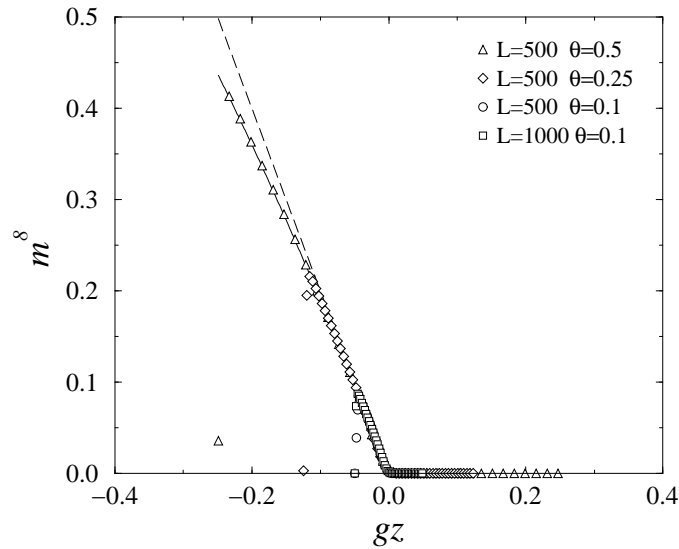}}
        \end{center}
	\vskip -.8cm
        \caption{Behaviour of the magnetization profile for $z<0$. The dashed line gives  $m^8=-2gz$, the behaviour 
	  expected from  equation~\eref{2-10}, valid in the scaling limit where $g\to0$. The solid line corresponds to
	   $m^8=-2gz -g^2z^2$ (see text). 
	 The values of the gradient are the same as in figure~\ref{Fig1}.}
        \label{Fig3}  
\end{figure}
%%%%%%%%%%%%%%%%%%%%%%%%%%%%%%%%%%%%%%%%%%%%%%%%%%%
%%%%%%%%%%%%%%%%%%%%%%%%% fig 4 %%%%%%%%%%%%%%%%%%%%%%%%
\begin{figure} [ht]
        \epsfxsize=9cm
        \begin{center}
        \mbox{\epsfbox{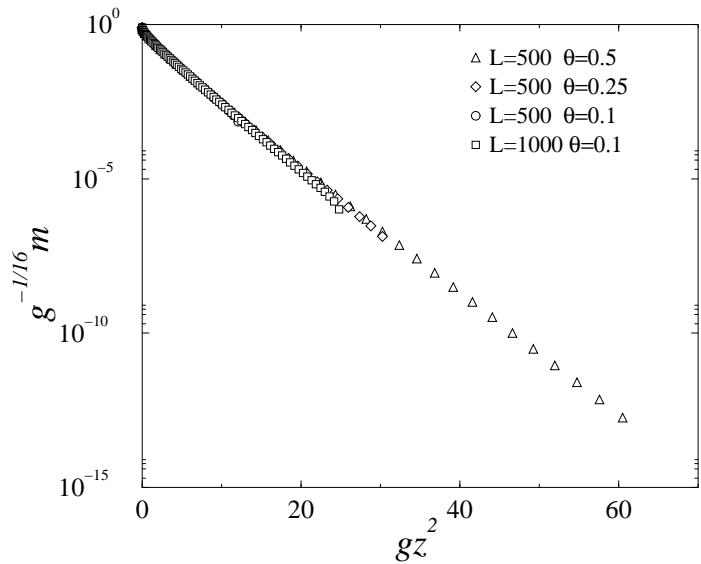}}
        \end{center}
	 \vskip -0.8cm
        \caption{Semi-logarithmic plot of the rescaled magnetization profile in the disordered region, $z>0$. 
	A linear behaviour is expected from equation~\eref{2-14}. The values of the gradient are
	 the same as in figure~\ref{Fig1}.}
        \label{Fig4}  
\end{figure}
%%%%%%%%%%%%%%%%%%%%%%%%%%%%%%%%%%%%%%%%%%%%%%%%%%%

Rewriting $\sigma_l^x$ in terms of diagonal fermions and using Wick's theorem, the local magnetization can be expressed as a determinant~\cite{berche96}
\be
\fl
m(l)=\left| \begin{array}{lllll}
H_{-L/2}&G_{-L/2,-L/2}&G_{-L/2,-L/2+1}& \dots&G_{-L/2,\,l-1}\\
H_{-L/2+1}&G_{-L/2+1,-L/2}&G_{-L/2+1,-L/2+1}& \dots&G_{-L/2+1,\,l-1}\\
\vdots&\vdots&\vdots&      &\vdots\\
H_{l-1}&G_{l-1,-L/2}&G_{l-1,-L/2+1}& \dots&G_{l-1,\,l-1}\\
H_l&G_{l,-L/2}&G_{l,-L/2+1}& \dots&G_{l,\,l-1}\end{array}
\right|
\label{4-28}
\ee
where
\be
H_{j}=\phi_{0}(j)\,,\qquad G_{j,k}=-\sum_{n=0}^L\phi_n(j)\psi_n(k)\,.
\label{4-29}
\ee
Figure~\ref{Fig2}  shows  the rescaled magnetization profile obtained for chain sizes up to $L=1000$ and different values of $\theta$.  The numerical results are in excellent agreement with the  scaling behaviour of equation~\eref{2-14} with $\nu=1$ and $x_{\rm m}=1/8$ for the Ising quantum chain in a transverse field. 

In order to test the scaling assumption~\eref{2-10}, $m^{1/(\nu x_{\rm m})}=m^8$ is plotted as a function of $gz$ in figure~\ref{Fig3}. The spontaneous  magnetization of the homogeneous Ising chain varies as $m=(1-h^2)^{1/8}$ as a function of the transverse field~\cite{pfeuty70} for $h\leq 1$. Replacing $h$ by its  local value $h(z)=1+gz$ in the inhomogeneous system, one obtains $[m(z,g)]^8=-2gz-g^2z^2$ for $z<0$ in very good agreement with the numerical data. One may notice that the second term in this expression does not fit with the scaling form  which follows from~\eref{2-9} but this term becomes negligible in the scaling limit where $g\to0$ as may be verified on figure~\ref{Fig3}.

Figure~\ref{Fig4} shows the behaviour of the rescaled magnetization profile in the disordered region on a linear-logarithmic scale in order to check the exponential decay, $\exp\left(-{\rm const.} g z^2\right)$,  expected  from equation~\eref{2-14}. The exact numerical results are once again in quite good agreement with the expected scaling behaviour.

\subsection{Correlation functions}

%%%%%%%%%%%%%%%%%%%%%%%%% fig 5 %%%%%%%%%%%%%%%%%%%%%%%%
 \begin{figure} [ht]
        \epsfxsize=9cm
        \begin{center}
        \mbox{\epsfbox{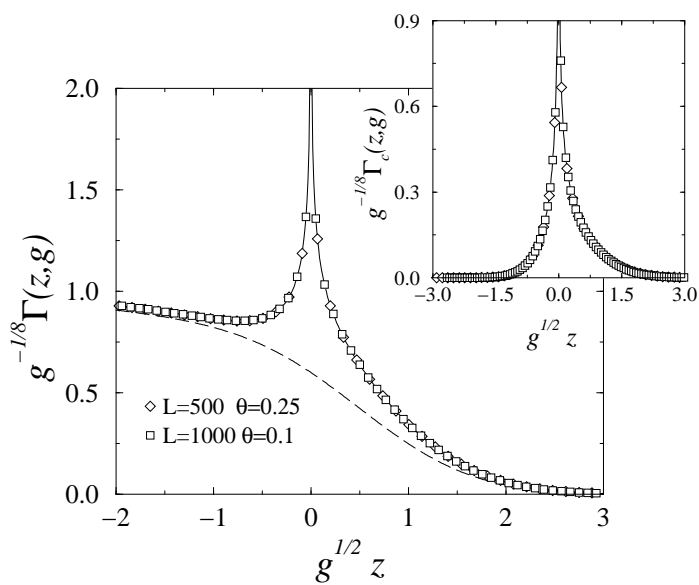}}
        \end{center}
	 \vskip -.8cm
        \caption{Scaling behaviour of the spin-spin correlation function $\Gamma(z,g)$ for $g=5\times 10^{-4}$
	 (diamond), $1\times 10^{-4}$  (square). The dashed line corresponds to the  product 
	 $ \langle m(0)\rangle\langle m(z)\rangle$ and the inset gives the behaviour of the  connected part.}
        \label{Fig5} 
\end{figure}
%%%%%%%%%%%%%%%%%%%%%%%%%%%%%%%%%%%%%%%%%%%%%%%%%%%%
%%%%%%%%%%%%%%%%%%%%%%%%% fig 6 %%%%%%%%%%%%%%%%%%%%%%%%
\begin{figure} [ht]
        \epsfxsize=9cm
        \begin{center}
        \mbox{\epsfbox{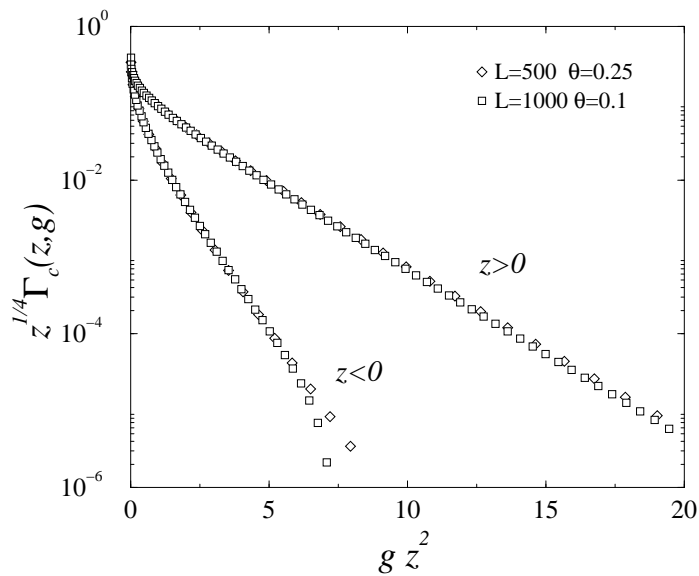}}
        \end{center}
	 \vskip -.8cm
        \caption{Semi-logarithmic plot of the rescaled connected part of the spin-spin correlation function 
	as a function of  $gz^2$. The linear behaviour  is in agreement with~\protect\eref{4-32}.}
        \label{Fig6} 
\end{figure}
%%%%%%%%%%%%%%%%%%%%%%%%%%%%%%%%%%%%%%%%%%%%%%%%%%%%

The correlation between a spin at $l$ and the central spin is measured by the correlation function
\be
\Gamma(l)=\langle 0|\sigma_0^x\sigma_l^x|0\rangle
\label{4-30}
\ee
given by the determinant~\cite{pfeuty70}
\be
\Gamma(l)=\left| \begin{array}{llll}
G_{1,0}&G_{1,1}& \dots&G_{1,l-1}\\
G_{2,0}&G_{2,1}& \dots&G_{2,l-1}\\
\vdots&\vdots      &      &\vdots\\
G_{l,0}&G_{l,1}& \dots&G_{l,l-1}\end{array}
\right|
\label{4-31}
\ee
which involves  the contractions  already defined in~\eref{4-29}.

The  numerical results,  shown in figure~\ref{Fig5}, confirm the scaling form given in~\eref{2-16}. The behaviour of the connected part  is shown  in the inset. A closer analysis of the  decay of $\Gamma_{\rm c}(z,g)$  points to  the following Gaussian behaviour 
\be
\Gamma_{\rm c}\sim z^{-1/4}\exp\left(-{\rm const.} g z^{2}\right) 
\label{4-32}
\ee
in agreement with the scaling prediction~\eref{2-16}.  The constant takes a different value on the two sides of the system (see figure~\ref{Fig6}).

\section{Summary and conclusion}
We have presented a scaling theory for the behaviour of the magnetization profile, the energy density profile and the two-point correlation function in a critical system in the presence of  a linearly varying deviation from the critical coupling, $\Delta(z)=gz$. The gradient $g$  introduces a new length scale in the problem, $\ell\sim g^{-\nu/(1+\nu)}$, which depends on the correlation length exponent $\nu$ in the case of a thermal perturbation considered here. The form  of the scaling functions have been obtained by assuming  that the different physical quantities have locally  the same functional form as in the homogeneous system with the deviation from the critical point, $\Delta$, replaced by its  local value, $\Delta(z)$, in the inhomogeneous system.  

The results of the scaling theory have been confirmed, first in mean-field theory and then  in  a study of the Ising quantum 
chain in a linearly varying transverse field. In this later case the excitation spectrum of the quantum Hamiltonian has been obtained exactly in the scaling limit where the size  of the system  $L\to\infty$, the gradient $g\to0$ while the product $Lg$ is held fixed. In this continuum limit one recovers the eigenvalue equation of the  harmonic oscillator problem. The knowledge of the eigenvectors allows us to calulate exactly the energy density profile and to obtain numerically exact results for the magnetization density and the two-point correlation function.  The very good agreement with the scaling results strongly supports the locality assumption used to deduce the form of the scaling functions.

\end{document}